\documentclass[pre,aps, notitlepage, balancelastpage, superscriptaddress, twocolumn]{revtex4-1}
\usepackage{mathptmx}
\usepackage{amssymb}
\usepackage{amsmath}
\usepackage[colorlinks,urlcolor=blue,linkcolor=blue,citecolor=blue]{hyperref}
\usepackage{mathtools}
\usepackage{mathrsfs}
\usepackage{graphicx}
\usepackage{amsfonts}
\usepackage{color}
\usepackage{bm}

\usepackage{xspace,soul}
\usepackage[]{hyperref}
\hypersetup{
    colorlinks,
    linkcolor={red!80!black},
    citecolor={blue!80!black},
    urlcolor={blue!80!black}
}
\usepackage{tikz}
\usepackage{braket}
\usepackage[]{algorithm2e}
\RestyleAlgo{ruled}
\usetikzlibrary{calc,decorations.markings}
\usepackage{comment}
\makeatletter

\def\clearemail{\let\@FMN@list\@empty}
\makeatother

\DeclarePairedDelimiter{\abs}{\lvert}{\rvert}
\DeclarePairedDelimiter{\norm}{\lVert}{\rVert}
\usepackage{braket}

\usepackage{tikz}
\usetikzlibrary{decorations.pathreplacing}
\usetikzlibrary{fadings}
\usepackage{pgffor}

\usepackage[normalem]{ulem}

\graphicspath{{images/}}

\newcommand{\btheta}{\boldsymbol{\theta}}
\newcommand{\bTheta}{\boldsymbol{\Theta}}

\newcommand{\bKappa}{\boldsymbol{K}}
\newcommand{\by}{\boldsymbol{y}}
\newcommand{\bkappa}{\boldsymbol{\kappa}}

\begin{document}

\title{Bayesian Optimization for QAOA}

\author{Simone Tibaldi}
\affiliation{Dipartimento di Fisica e Astronomia dell'Universit\`a di Bologna, I-40127 Bologna, Italy}
\affiliation{INFN, Sezione di Bologna, I-40127 Bologna, Italy}

\author{Davide Vodola}
\affiliation{Dipartimento di Fisica e Astronomia dell'Universit\`a di Bologna, I-40127 Bologna, Italy}
\affiliation{INFN, Sezione di Bologna, I-40127 Bologna, Italy}

\author{Edoardo Tignone}
\affiliation{Leithà S.r.l.~\text{\textbar} Unipol Group, Bologna, Italy}

\author{Elisa Ercolessi}
\affiliation{Dipartimento di Fisica e Astronomia dell'Universit\`a di Bologna, I-40127 Bologna, Italy}
\affiliation{INFN, Sezione di Bologna, I-40127 Bologna, Italy}

\begin{abstract}
The Quantum Approximate Optimization Algorithm (QAOA) adopts a hybrid quantum-classical approach to find approximate solutions to variational optimization problems. In fact, it relies on a classical subroutine to optimize the parameters of a quantum circuit. In this work we present a Bayesian optimization procedure to fulfil this optimization task, and we investigate its performance in comparison with other global optimizers. We show that our approach allows for a significant reduction in the number of calls to the quantum circuit, which is typically the most expensive part of the QAOA. We demonstrate that our method works well also in the regime of slow circuit repetition rates, and that few measurements of the quantum ansatz would already suffice to achieve a good estimate of the energy. In addition, we study the performance of our method in the presence of noise at gate level, and we find that for low circuit depths it is robust against noise. Our results suggest that the method proposed here is a promising framework to leverage the hybrid nature of QAOA on the noisy intermediate-scale quantum devices.
\end{abstract}
\maketitle	

\section{Introduction}
Hybrid quantum-classical variational algorithms~\cite{McClean:2016uo, Peruzzo:2014uz, Moll:2018vf} play a central role in the current research on noisy intermediate-scale quantum (NISQ) devices~\cite{Preskill_2018}. In a hybrid variational setting, a classical computer is entrusted with the non-trivial task of optimizing the parameters of a quantum state. These algorithms implement a heuristic protocol to approximately solve variational problems including combinatorial optimization tasks, which are ubiquitous and have a great practical importance \cite{DP1999}, and are indeed one of the main drivers of the industry interest towards quantum computing applications. Unfortunately, the problems belonging to this class are hard to solve with classical methods \cite{GareyJ79}. In this paper we focus on the Max-Cut and the Max Independent Set (MIS) problems defined on specific graph instances.  

Among the hybrid variational algorithms, the Quantum Approximate Optimization Algorithm (QAOA)~\cite{Farhi:2014wk} is extensively studied \cite{Zhou:2020th} as a promising algorithm to investigate quantum speedups on NISQ devices and has been implemented on several experimental platforms, such as Rydberg atom arrays \cite{Ebadi2022}, superconducting processors \cite{Harrigan:2021we}, trapped-ions simulators \cite{Pagano2020}, as well as simulated on classical devices \cite{Medvidovic:2021aa}.

Similarly to other hybrid variational algorithms, QAOA consists in a sequence of parametrized quantum gates applied to a wavefunction, on which an expectation value of some operator, typically the Hamiltonian, is reconstructed from measurements. The task of the classical subroutine is to optimize the gate parameters in order to minimize such expectation value. Every variational quantum algorithm therefore requires the estimation of the wavefunction, which is a sub-optimal problem that requires resources that grow exponentially with the system size~\cite{Cramer:2010uo}. Moreover, the interplay between the classical and quantum parts of the algorithm entails to run the quantum circuit a large number of times, thus being expensive in term of resources. Finally, there is the notorious problem of Barren Plateaus (BP) which are large portions of the optimization landscape in which the gradient becomes exponentially small with the number of qubits and layers ~\cite{McClean2018}. This phenomenon was proven to be caused also by the presence of noise~\cite{Wang2021} or by the use of a cost function depending on global observables~\cite{Cerezo2021}. 

To overcome these issues, in fact, an efficient classical optimization routine is crucial. Different techniques have been proposed for optimizing variational quantum circuits, e.g. Nelder-Mead~\cite{Guerreschi:2017vm}, Machine Learning~\cite{Wecker:2016wz}, gradient descent~\cite{Wang:2018vs}, iterative schemes~\cite{Zhou:2020th, Mele:2022rqi, Mbeng2019QuantumAA}, Gaussian Processes~\cite{Shaffer2023,Mueller2022} and Bayesian methods~\cite{Otterbach2017,Zhu2019,Self2021,Wang2021,Tamiya2022}. In particular, to tackle the problem of BPs it might seem logical to avoid the calculation of the gradient. However, in~\cite{Arrasmith2021} it was shown that gradient-free optimizers such as COBYLA, Powell and Nelder-Mead suffer from BPs too. Here, we focus on a Bayesian optimization framework, which is suitable for gradient-free global optimization of black-box functions~\cite{Shahriari2016, Frazier2018}. We explore its behaviour in comparison with other global optimizers and we show that the convergence rate to a minimum is faster. We demonstrate that the Bayesian approach is efficient in terms of number of circuit runs and is robust against noise sources. 

The paper is structured as follows. In Section~\ref{QAOA} we introduce the QAOA algorithm. In Section~\ref{sec_BayesianOptimization} we give a detailed presentation of the Bayesian algorithm. In Section~\ref{Results} we present the result of applying this method to QAOA, compare it to other global optimization methods, evaluate its performance with a limited number of circuit runs and against simulated quantum noise. Finally, in Section~\ref{Conclusions} we draw our conclusions.

\section{QAOA for combinatorial problems}\label{QAOA}
The QAOA is a variational quantum algorithm that performs hybrid quantum-classical optimization~\cite{Farhi:2014wk}. 
Given a cost function $C(z)$ with $z=(z_1, \dots, z_i, \dots, z_N)$ with $ z_i \in \{0,1\}$, QAOA aims at finding the bitstring $z^\star$ that minimizes the cost. In order to do so, the cost function is translated into a quantum operator $H_C$ that is diagonal in the computational basis $\ket{z} = \ket{z_1 \dots z_i \dots z_N}$ for $N$ qubits i.e. $H_C \ket{z} = C(z) \ket{z}$.

The QAOA circuit consists in preparing an initial state of $N$ qubits, usually $|+ \rangle = \sum_z |z \rangle / \sqrt{2^N}$, and then applying two unitary operators alternatively: one generated by $H_C$, the other generated by $H_M = \sum_i X_i$, where $X_i$ is the flip (NOT) operator on the $i$-th qubit for a number of layers $p$ which is called the depth of the circuit.
In this way the QAOA circuit prepares the state
\begin{equation}
	\label{eq:qaoa_state}
	|\boldsymbol{\theta}\rangle = \prod_{l=1}^p e^{-i\beta_l H_M} e^{-i \gamma_l H_C} |+  \rangle,
\end{equation}
where $\boldsymbol{\theta} = (\boldsymbol{\gamma},\boldsymbol{\beta})$ are $2p$ parameters. By measuring the state $\ket{\boldsymbol{\theta}}$ in the computational basis an estimate of the energy $E(\boldsymbol{\theta}) = \langle \boldsymbol{\theta} | H_C | \boldsymbol{\theta} \rangle$ is obtained. This energy is fed to a classical routine which looks for the set of angles $\boldsymbol{\theta}^\star = (\boldsymbol{\gamma}^\star, \boldsymbol{\beta}^\star)$ that minimizes $E(\boldsymbol{\theta})$.
Several strategies have been proposed for finding the optimal parameters $\boldsymbol{\theta}^\star$. In this work we rely on Bayesian optimization.

\section{Bayesian optimization}\label{sec_BayesianOptimization} 
Bayesian optimization is a global optimization strategy which allows to find within relatively few evaluations the minimum of a noisy, black-box objective function $f(\btheta)$ that is in general expensive to evaluate~\cite{Snoek2012}. 
The algorithm can be summarized as follows: (i) it treats the objective function $f$ as a random function by choosing a prior (also called surrogate model) for $f$. Several choices for the surrogate model are possible~\cite{Shahriari2016}, in this work we adopt the so-called Gaussian process~\cite{Rasmussen2005}. (ii)  The prior is then updated through the likelihood function by gathering observations of~$f$ and therefore forming the posterior distribution.
(iii) The posterior distribution is finally used to construct an auxiliary function, called acquisition function, that is in general cheap to evaluate. The point where the acquisition function is maximized gives the next point where~$f$ will be evaluated~\cite{Frazier2018}. See Appendix~\ref{sec_details_BO_1} for an overview of Bayesian terminology.  

Since Bayesian optimization requires no previous knowledge on $f$, it appears to be a well-suited technique for optimizing the parameters of a variational circuit running on NISQ devices.

In the following sections we describe the Gaussian process, the optimization routine and the acquisition function in detail.

\subsection{Gaussian process}

Since the  function $f(\btheta)$  ($\btheta \in A \subset \mathbb{R}^d$) to be optimized is unknown, we may think of it as belonging to a random process, i.e.~an infinite collection of random variables defined for every point $\btheta \in A$. A random process is called Gaussian if the joint distribution of any finite collection of those random variables is a multivariate normal distribution defined by a mean function $	\mu(\btheta)$ and covariance (or kernel) function $k(\btheta, \btheta')$~\cite{Rasmussen2005}. The mean function is related to the expected value of the function $f$, while the kernel estimates the  deviations of the mean function from the value of $f$:
\begin{eqnarray}
	\mu(\btheta) &=& \mathbb{E}[f(\btheta)],\label{eqn_mean_function}  \\
	k(\btheta, \btheta') &=& \mathbb{E}[(f(\btheta) - \mu(\btheta))(f(\btheta') - \mu(\btheta'))],\label{eqn_cov_function} 
\end{eqnarray}
where $\mathbb{E}$ denotes the expectation w.r.t.~the infinite collection of functions belonging to the random process. Conceptually, the mean encloses the knowledge of the function $f$ to reconstruct while $k$ represents the uncertainty we have on such reconstruction.

Since we assume $f$ to be smooth, we choose for $k$ the Mat\'ern kernel, a stationary kernel~\cite{Rasmussen2005} that depends on the distance between the points $\btheta$ and $\btheta'$, defined as
\begin{equation}
	\label{matern_kernel}
	 k(\btheta, \btheta') = \sigma^2 \bigg( 1 + \frac{\sqrt{3} \norm{\btheta - \btheta'}_2}{\ell} \bigg) e^{-\frac{\sqrt{3} \norm{\btheta - \btheta'}_2} {\ell}},
\end{equation}
where $\norm{\cdot}_2$ is the 2-norm and $\sigma^2$ and $\ell$ are two hyperparameters characterizing the Gaussian process. The hyperparameter $\sigma^2$ defines the variance of the random variables whereas $\ell$ is a characteristic length-scale which regulates the decay of the correlation between points: in the limit of $\ell \rightarrow \infty$ all points are equally correlated, for $\ell \rightarrow 0$ all points are uncorrelated.

\subsection{Bayesian optimization algorithm}

The main steps of the algorithm for Bayesian optimization can be summarized in the pseudocode in Algorithm~\ref{alg_pseudocode} (see also Appendix~\ref{sec_details_BO_2} for details).

\begin{algorithm}[h]
\caption{Pseudo-code for Bayesian optimization}\label{alg_pseudocode}
Set the prior on $f$ as a Gaussian process\;
Evaluate $f$ at $N_W$ different points $\btheta_i$\;
Define the initial training set  $\mathcal{D} = \{(\btheta_i, f(\theta_i))\}_{i=1}^{N_W}$\;
Compute the hyperparameters $\sigma^2, \ell$ based on $\mathcal{D} $\;
Set the guess for the minimum of $f$ to $f_m = \min [\{f(\theta_i)\}_{i=1}^{N_W}$]\;
\While{$n \leq N_\mathrm{BAYES}$}{
	Update the posterior distribution on $f$ using the training set $\mathcal{D}$\;
	Compute the acquisition function with the updated posterior\;
	Find $\tilde{\btheta}$ that maximizes the acquisition function\;
	Evaluate $f(\tilde{\btheta})$\;
	\If{$f(\tilde{\btheta}) < f_m$}{Set the guess for the minimum of $f$ to $f_m =  f(\tilde{\btheta})$\;}
	Append $(\tilde{\btheta}, f(\tilde{\btheta}))$ to $\mathcal{D}$\;
	Compute the new hyperparameters $\sigma^2, \ell$\;
	Increment $n$\;
    }
\textbf{return} $f_m$
\end{algorithm}

The optimization starts by a warmup phase where a number $N_W$ of evaluations of the objective function $f$ is performed. These evaluations take place at  randomly chosen values of the points $\btheta_i$ and are collected in the training set $\mathcal{D} = \{(\btheta_i, y_i=f(\theta_i))\}_{i=1}^{N_W}$ of the optimization. Given the set $\mathcal{D}$, we define the design matrix $\bTheta = (\btheta_1,\dots,\btheta_{N_W})$ with the points and the vector $\by \in \mathbb{R}^{N_W}$ with the observations via $\by = (y_1,\dots, y_{N_W})$. We form the covariance matrix $\boldsymbol{K} \in \mathbb{R}^{N_W \times N_W}$ by evaluating the covariance function in Eq.~\eqref{eqn_cov_function} for each pair of points $\btheta_i, \btheta_j \in \bTheta$ via
\begin{equation}
\boldsymbol{K}_{i,j} = k(\btheta_i, \btheta_j),
\end{equation}
where $\bKappa_{i,j}$ denotes the $(i,j)$ element of the matrix $\boldsymbol{K}$. The hyperparameters entering the kernel function (Eq.\eqref{matern_kernel}) are optimized at this step, as explained in Sec.~\ref{sec_hyperparam}.

The training set will be used at each step of the optimization to incorporate the acquired knowledge in the Gaussian process. This happens in two steps.
First, the Gaussian process prior is conditioned on the observations in $\mathcal{D}$~\cite{Rasmussen2005}. Conditioning is equivalent to a Bayesian step in which we multiply the prior with the likelihood, thus obtaining a posterior distribution (see Appendix~\ref{sec_details_BO_1}). Thanks to the properties of Gaussian distributions, the posterior is still described by a Gaussian process multinomial distribution but it is characterized by a posterior mean $\mu'$ and covariance $k'$ given by
\begin{align}
		\mu' &= \bkappa^T \cdot \bKappa^{-1}\cdot\by \label{eqn_new_mean}\\
		k' &= k(\btheta, \btheta) - \bkappa^T  \cdot \bKappa^{-1}\cdot \bkappa.	\label{eqn_new_cov}
\end{align}
Here, $\btheta$ is a generic point in $A$ and $\bkappa$ is a column vector formed by evaluating the covariance function $k$ between the generic point $\btheta$ and all the points in $\bTheta$, i.e.~its $j$-th element is $\kappa_j = k(\btheta, \btheta_j)$. Eq.~\eqref{eqn_new_mean} shows that the new mean is a linear combination of the observations~$\by$.

\subsection{Acquisition Function}
The next step in the Bayesian optimization is computing the acquisition function, whose maximum gives the next point at which to evaluate the objective function. A common choice of acquisition function is the Expected Improvement (EI): this function suggests which points, on average, improve on $f_m$ the most~\cite{Frazier2018}. This choice corresponds to defining the acquisition function $\textrm{EI}(\btheta) = \mathbb{E}[u(\btheta)]$ as the average value of the utility function $u(\btheta) = \max[0, f_m - f(\btheta)]$ such that the lower $f(\btheta)$ is with respect to the current minimum, the larger the utility $u(\btheta)$ will be. 

By considering that $f(\btheta)$ is a Gaussian process, we can obtain an analytical expression for $\mathrm{EI}(\btheta)$ as
\begin{equation}
\mathrm{EI}(\btheta) =  \Phi (z)  (f_m - \mu') + \phi(z) k',  \label{eqn_acquisition_function}
\end{equation} 
where $\mu'$ and $k'$ are obtained for the point $\btheta$ by using Eqs.~\eqref{eqn_new_mean} and~\eqref{eqn_new_cov}; $\Phi(\cdot)$  and $\phi(\cdot)$ are respectively the cumulative distribution function and the probability density function of the standard normal distribution and the quantity $z$ is defined as $z = (f_m - \mu') / k'$. The two terms in Eq.~\eqref{eqn_acquisition_function} resume the trade-off between exploitation and exploration: the first term, being proportional to the difference between the current minimum and the mean value of the posterior, brings the optimization towards points with lower $\mu'$ whereas the second one promotes points with larger $k'$, i.e.~with higher uncertainty. The point $\tilde{\btheta}$ that maximizes the acquisition function is then added to the training set $\mathcal{D}$ and the algorithm's loop is repeated (as written in Algorithm~\ref{alg_pseudocode}). Its value is found by using the differential evolution algorithm~\cite{Price2006}, a population-based metaheuristic search algorithm (see Appendix~\ref{sec_diff_ev} for details).

\begin{figure}[b!]
	\includegraphics[width = \linewidth]{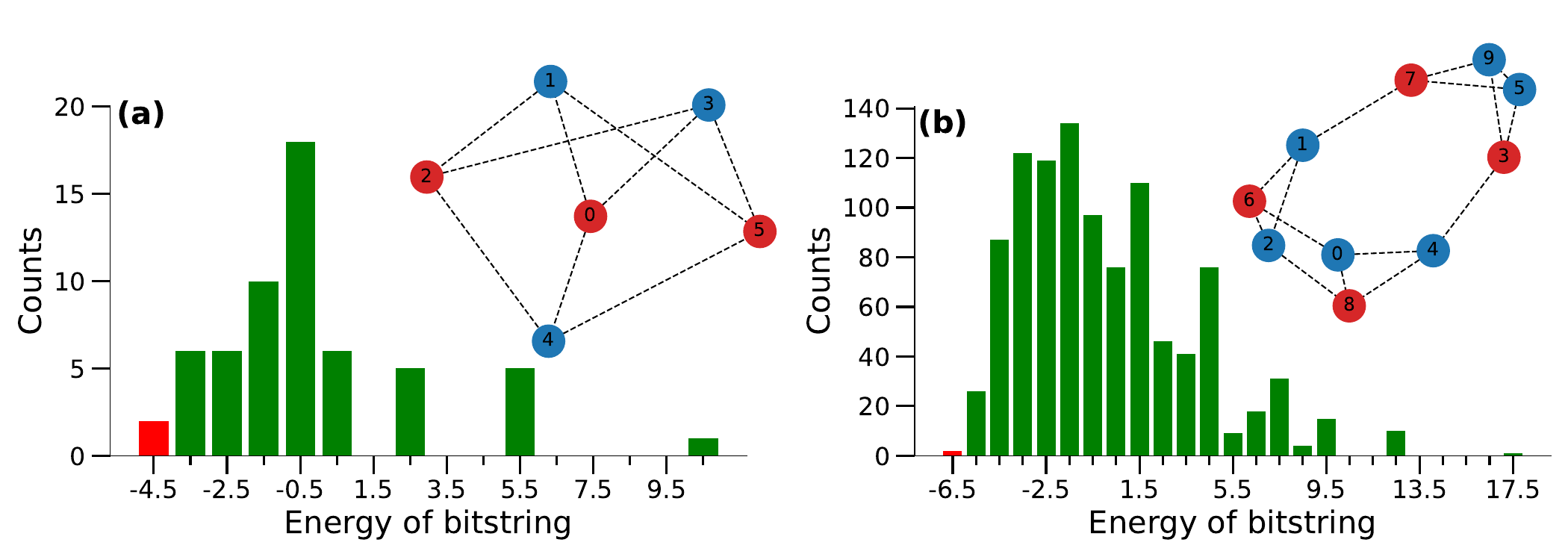}
	
	\caption{\textbf{Energy distributions of graphs.} (a) MIS: Distribution of the energies of the possible bitstrings for the graph of 6 nodes (shown in the inset, nodes in red correspond to one solution). The red bar to the left highlights the two solution bitstrings of the MIS problem on such graph. (b) Max-Cut: Distribution of the energies of the possible bitstrings for the graph of 10 nodes (shown in the inset, nodes in red correspond to one solution). The red bar highlights the two solution bitstrings of the Max-Cut problem on such graph.}
	\label{fig:graphs}
	
\end{figure}

\subsection{Hyperparameters}\label{sec_hyperparam}
We are now only left with the task of picking the best hyperparameters $\sigma, \ell$ for the Mat\'ern kernel. This is typically done by considering the marginal likelihood~\cite{Rasmussen2005} (and Appendix~\ref{sec_details_BO_1})
\begin{equation}
	p(\boldsymbol{y} | \bTheta) = \int p(\boldsymbol{y} | \boldsymbol{f}, \bTheta) p(\boldsymbol{f} | \bTheta) d\boldsymbol{f},
\end{equation}
where the prior $p(\boldsymbol{f} | \bTheta)$ and the likelihood $p(\boldsymbol{y} | \boldsymbol{f}, \bTheta)$ are Gaussian and the marginalization is done over the function values $\boldsymbol{f}$. Given the Gaussian nature of the likelihood and the prior,  a closed form of the log marginal likelihood can be obtained (for the standard derivation of this formula see for example~\cite{Rasmussen2005}):
\begin{align}
	\begin{split}
	\log p(\boldsymbol{y} | \bTheta) = &-\frac{1}{2} \boldsymbol{y}^T \cdot \bKappa^{-1} \cdot \boldsymbol{y} \\
	&-\frac{1}{2} \log \det\bKappa - \frac{N}{2} \log 2 \pi,
	\end{split}
	\label{eqn_log_marginal_likelihood}
\end{align}
where $N$ is the number of observations in the design matrix $\bTheta$. In Eq.~\eqref{eqn_log_marginal_likelihood}, the first term specifies how well the process fits the data, the second term instead acts as a regularization factor on the elements of the kernel matrix. 
When fitting the Gaussian process to a new set of points, the best hyperparameters $(\tilde{\sigma}^{2}, \tilde{\ell})$ can be found by maximizing the log marginal likelihood in Eq.~\eqref{eqn_log_marginal_likelihood}. For the optimization of $\log p(\boldsymbol{y} | \bTheta) $, we use the quasi-Newton method \texttt{L-BFGS}~\cite{Liu1989} with multiple restarting points which proved to be efficient on the flat landscape of the likelihood (see Appendix~\ref{sec_details_BO_2} for details).

\section{Results}\label{Results}
In this section we apply the Bayesian optimization to the QAOA parameters. We consider two well-known combinatorial problems defined on graphs, the Max-Cut and the Max Independent Set.

\textit{Max-Cut --} 
Given a graph $G = (V, E)$ where $V$ is the set of nodes and $E$ the set of edges, the Max-Cut problem consists in finding a partition of the graph's vertices $V$, $P=\{V_0, V_1\}$, such that the number of edges between $V_0$ and its complement $V_1$ is as large as possible. It is known to be a NP-hard problem~\cite{Edwards:1973ux}. We can define the assignment of the nodes to the sets $V_0$ and $V_1$ by labelling with label ``0'' the nodes $v \in V_0$ and with label ``1'' the nodes $v \in V_1$. In these terms, the Max-Cut consists in finding the largest number of edges connecting the bits labelled with ``0'' to the bits labelled with ``1''. On a quantum computer, the labels 0 and 1 are replaced by the computational basis states $\ket{0}$ and $\ket{1}$ and the cost Hamiltonian can be written as:
\begin{equation}
\label{eq:maxcut_hamiltonian}
	H_C^\text{MC} = -\sum_{(i,j) \in E} (1 - Z_i Z_j)/2.
\end{equation}
The cost function has minimum eigenvalue on the states $\ket{z_i z_j}$ for which $z_i \neq z_j$ which represent separate partitions.

\begin{figure}[b!]
	\centering
	\includegraphics[width = 0.8\linewidth]{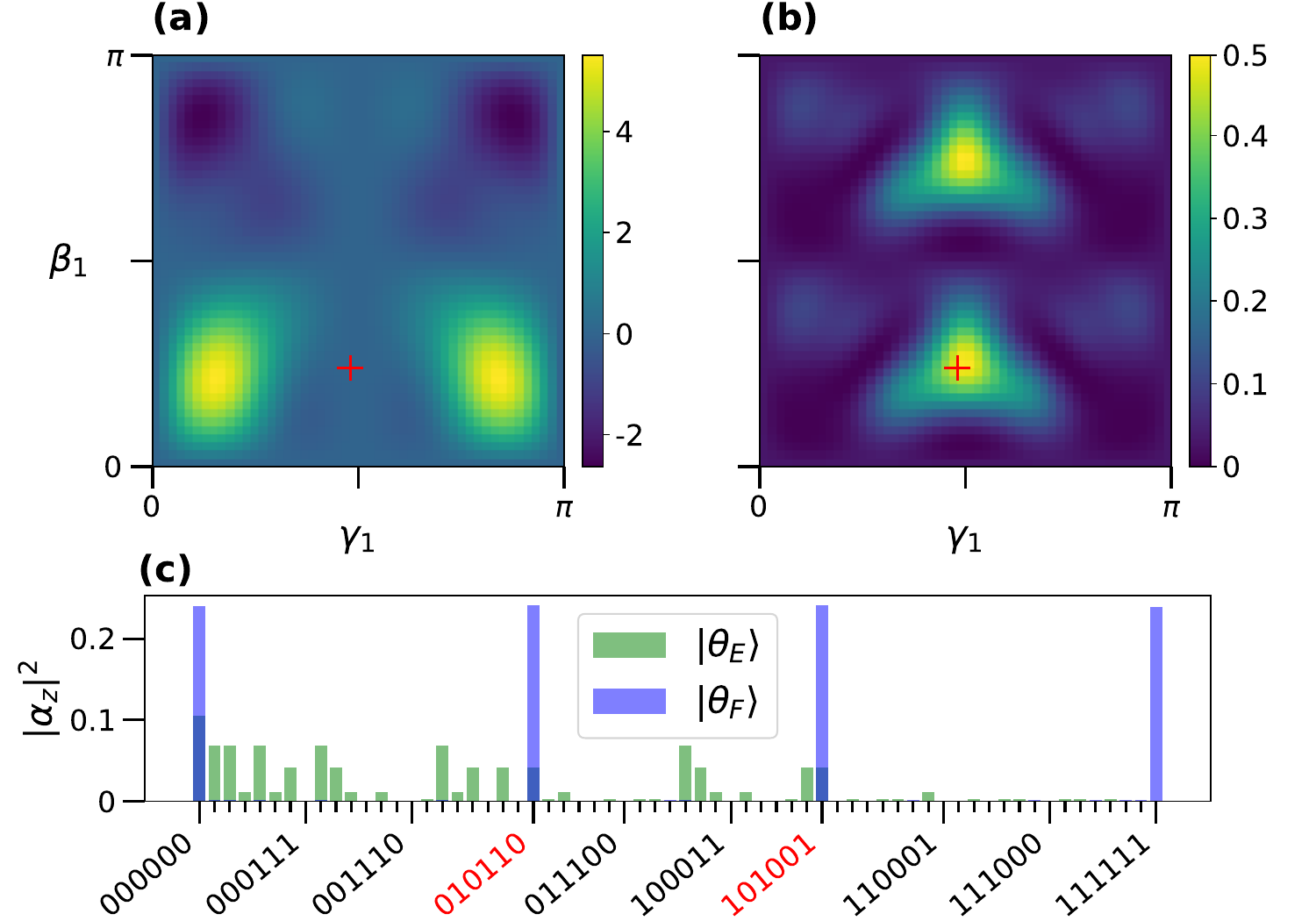}
	
	\caption{\textbf{QAOA at $\boldsymbol{p = 1}$.} (a) Landscape of the energy $E(\boldsymbol{\theta}_1)$ obtained on the 6 nodes graph solving the MIS problem. The red cross indicates the angles corresponding to the final state $\ket{\boldsymbol{\theta}_F}$ with largest fidelity. (b) Landscape of the fidelity $F(\boldsymbol{\theta}_1)$. (c) Squared amplitudes of the two states $\ket{\boldsymbol{\theta}_E}$, $\ket{\boldsymbol{\theta}_F}$. The solution bitstrings are highlighted in red. Values of the qubits are given in the order shown in the inset of Fig.~\ref{fig:graphs}(a).}
	\label{fig:landscapes}
\end{figure}

\textit{MIS --} 
The Max Independent Set problem consists in finding the largest number of graph nodes which are not adjacent and the corresponding cost quantum Hamiltonian is:
\begin{equation}
\label{eq:mis_hamiltonian}
	H_C^\text{MIS} = \sum_{i} Z_i + \omega \sum_{(i,j) \in E} Z_i Z_j,
\end{equation}
with $\omega$ a parameter that balances the effect of the first term (which maximizes the number of qubits in $\ket{1}$) and the second one (which prevents neighbour bits to be equal). We set $\omega = 2$ for the rest of the work.

During the optimization process we monitor the approximation ratio $R = E(\boldsymbol{\theta})/E_{GS}$ ~\cite{Farhi:2014wk} where $E_{GS}$ is the energy of the solution bitstring. Since $E_{GS} < 0$ (due to our definition of the problems's hamiltonians~\eqref{eq:maxcut_hamiltonian},~\eqref{eq:mis_hamiltonian}), $|E(\boldsymbol{\theta})| \leq |E_{GS}|$ and thus $R$ is upper bounded by 1. We also look at the fidelity defined as $F = |\braket{\boldsymbol{\theta}|z^\star} |^2$  where $\ket{z^\star}$ is the state which encodes the solution. The following results are obtained on two 3-regular graphs of 6 and 10 nodes which are plotted in the insets of Figs.~\ref{fig:graphs}(a)--(b).

\textit{QAOA at low vs.~large depth --} We start by looking at the QAOA at depth $p=1$ for the 6 nodes graph. It corresponds to a shallow circuit that depends only on two parameters $\boldsymbol{\theta_1} = (\gamma_1, \beta_1)$. We consider the MIS problem, and we plot the landscapes of both the energy $E(\boldsymbol{\theta}_1)$ and the fidelity $F(\boldsymbol{\theta}_1)$ (Fig.~\ref{fig:landscapes}(a) and (b), respectively) for values of $\gamma_1, \beta_1 \in [0, \pi]$ due to the symmetry of the problem. We see that the landscape of the energy, which is the function to minimize, is rather flat with two global maxima and minima, corresponding to the best solutions possible at $p=1$. 

Interestingly, we find that the QAOA state $\ket{\boldsymbol{\theta_E}} = \sum_z \alpha_{z,E} \ket{z}$, corresponding to the parameters which minimize the energy, is not the state $\ket{\boldsymbol{\theta_F}}  = \sum_z \alpha_{z, F} \ket{z}$ with largest fidelity. To see how they differ we plot the squared amplitudes $\abs{\alpha_{z, E}}^2$ and $\abs{\alpha_{z, F}}^2$ of both states in Fig.~\ref{fig:landscapes}(c) as histograms. The fidelity of $\ket{\boldsymbol{\theta_F}}$ w.r.t.~the solution $\ket{z^\star}$ is, as expected, much larger than that one of $\ket{\boldsymbol{\theta_E}}$, yet the latter has a lower energy because it has many non zero amplitudes along excited states with low energy. This unravels the problem of optimizing the QAOA parameters by only looking at the energy $E(\boldsymbol{\theta})$. There is a large concentration of excited states with energy comparable to the energy of the ground state, as shown in the histograms of panels (a) and (b) of Fig.~\ref{fig:graphs}. It is difficult to increase the amplitude corresponding to the solution when many other states can contribute with low values of the energy.

\begin{figure}
	\centering
	\includegraphics[width = 0.5\linewidth]{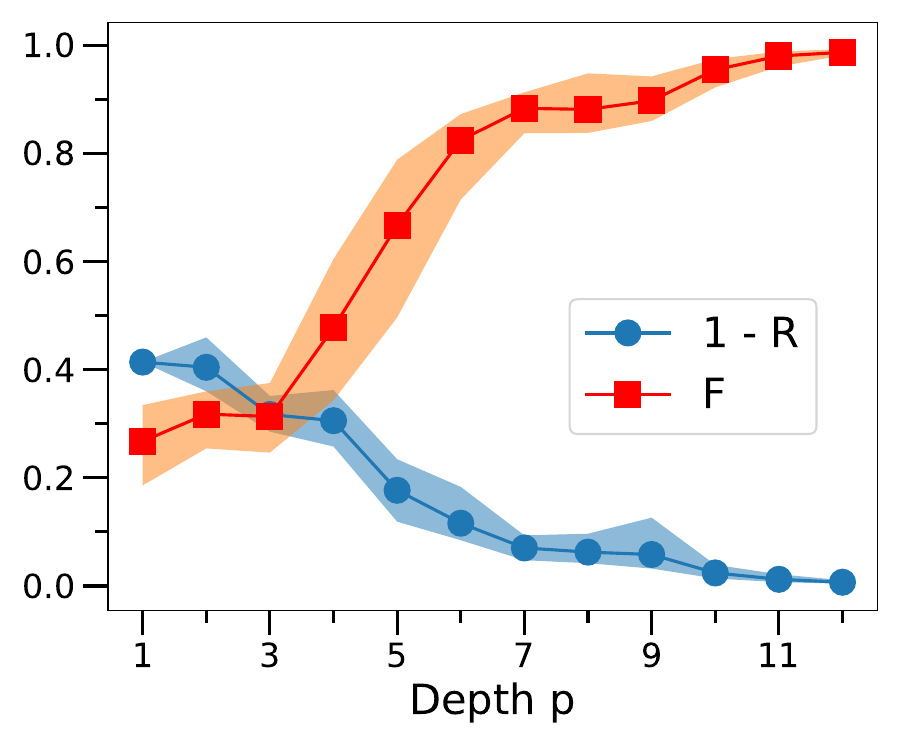}
	
	\caption{\textbf{Results increasing depth.} Average approximation ratio (plotted as $1 - R$) and fidelity $F$ for increasing values of circuit depth from 1 to 12 over 50 runs. Shaded areas correspond to one standard deviation. Results were obtained on the 6 nodes graph of Fig.~\ref{fig:graphs}(a).
}
	\label{fig:MIS}
	
\end{figure}

The divergence between lowest energy and highest fidelity is guaranteed to disappear theoretically for $p \rightarrow \infty$. For this reason, we apply Bayesian optimization to the problem and we show in Fig.~\ref{fig:MIS} that the approximation ratio and fidelity both tend to 1 for $p \sim 12$. Yet, we already see a good performance at $p=4$ where $R \sim 0.7$ and we have $ F \sim 0.5$ meaning about a $50\%$ chance of measuring the solution on the state obtained with QAOA.

\begin{figure*}
	\centering
	\includegraphics[width = \linewidth]{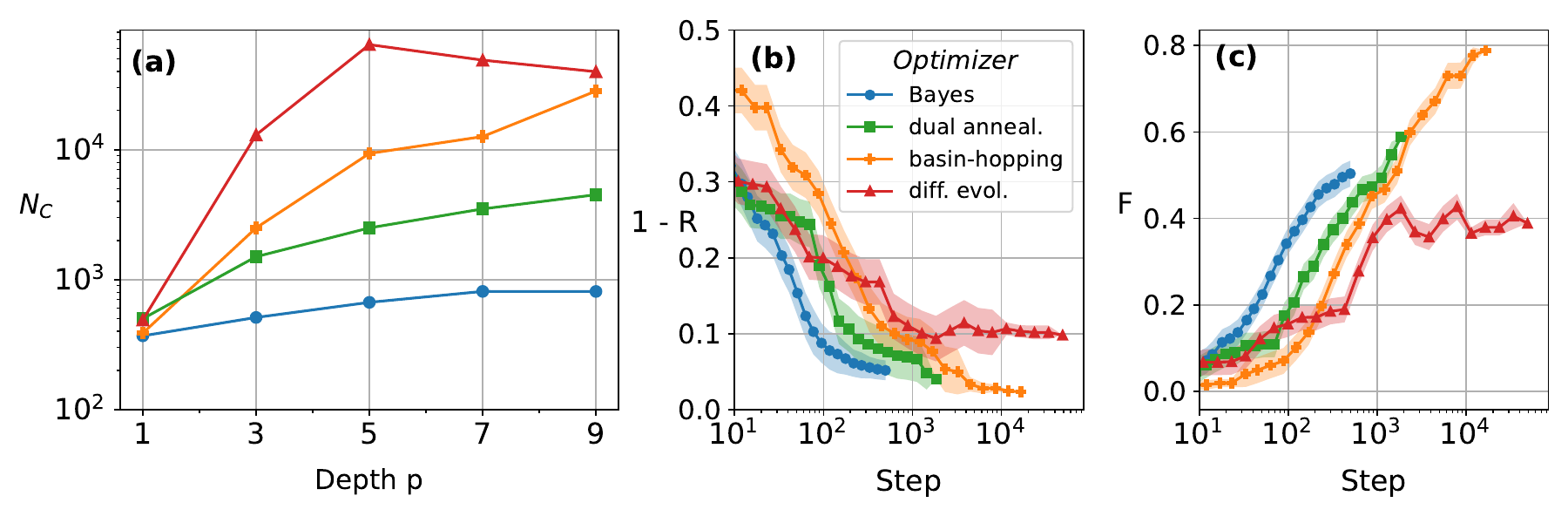}
	\caption{\textbf{Comparison among optimizers}. (a) The plot shows the average number of calls $N_C$ to the quantum circuit  of each optimizer in order to obtain the same approximation ratio as the Bayesian optimization. (b) - (c) Average approximation ratio (plotted as $1 - R$) and fidelity during the optimization with the different methods at $p = 7$ over 30 runs. Shaded areas correspond to one standard deviation. Results are obtained on the 10 nodes graph of Fig.~\ref{fig:graphs}(b).
}
	\label{fig:comparison}
\end{figure*}

\textit{Comparing resources -- } 
Increasing the depth of a variational circuit increases the number of parameters that must be optimized. In turn, this is expected to increase the number of calls to the quantum circuit needed to reach a good approximate solution, which is a problem in the current NISQ era, since running a quantum circuit can be costly due to both state preparation routines and recalibrations of the device. 

Bayesian optimization mitigates such problem, and allows to achieve a good approximate solution within relatively fewer calls to the circuit compared to other global optimization methods. To show this, we ran differential evolution, basin-hopping and dual annealing (see Appendix~\ref{sec_diff_ev} and \ref{other_optimizers} for details) on the 10 nodes graph of Fig.~\ref{fig:graphs}(b) for the Max-Cut problem at different depths. Fig.~\ref{fig:comparison}(a) shows the average number of calls to the circuit that the other optimizers need in order to reach the same approximation ratio of Bayesian optimization. To gain more insight, we plot in panels (b) and (c) of Fig.~\ref{fig:comparison} the average approximation ratio and fidelity during the run of the algorithm for each method at $p=7$. We see that Bayesian optimization stops at a lower $R$ than basin-hopping and dual annealing, but it reaches a value of $R = 95\%$ with $\sim 500$ runs of the circuit compared to the other two methods which take, in order, 1400 and 10800.

For the noiseless circuit, (it) is very clear (see Figure 4) that the Bayesian approach can mitigate this problem better than any other tested techniques, from different points of view (such as number of calls, number of steps, number of measurements). Let us stress that this approach gives a fidelity that is always higher that the other methods at a fixed number of steps (and fixed p, see Figure 4c). 

\textit{Slow Measurements --}
The energy $E(\boldsymbol{\theta})$ is obtained by measuring the QAOA state after running the circuit: we refer to these two operations combined together as a ``shot''. By measuring on the $Z$ basis at each shot we get a bitstring, and we calculate its classical energy associated to the combinatorial problem. The precision in the reconstruction of $E(\boldsymbol{\theta})$ depends on the number of shots $N_{S}$. Since we consider this as a multinomial sampling problem we expect the variance of the reconstructed energy to depend on $N_S^{-1}$. In many scenarios of NISQ devices it is necessary to balance $N_S$ with the desired standard deviation. For this reason, we compare the average approximation ratio obtained with the exact energy (simulated) with the energy reconstructed with a limited number of shots.

\begin{figure*}
	\includegraphics[width = \linewidth]{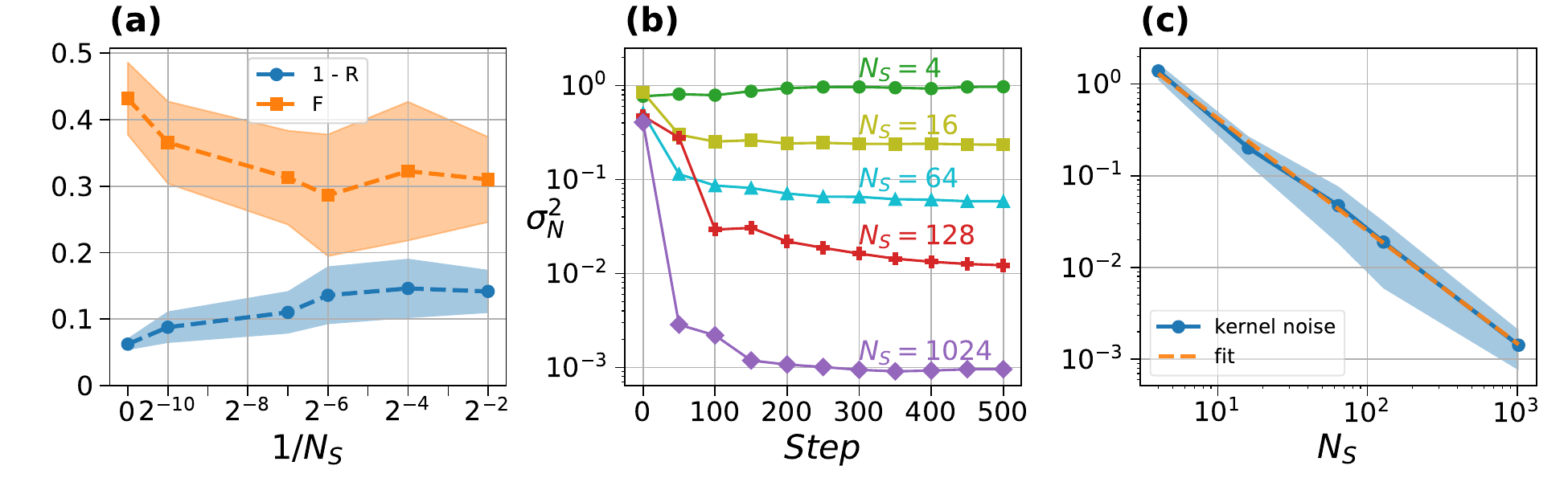}
	\caption{\textbf{Slow measurements.} (a) Average approximation ratio (plotted as $1 - R$) and fidelity~($F$) as a function of the number of shots $N_S$. The $1/N_S = 0$ points indicates the exact evaluation of the energy $E(\btheta)$. Shaded areas correspond to one standard deviation. (b) Kernel noise $\sigma^2_N$ learned by fitting the data at each step of the optimization for different number of shots $N_S$. (c) Average kernel noise learned by the Gaussian process as a function of $N_S$ (blue circles). The plot also shows a linear fit ($\sim 1/N^{1.1}$, orange line) of the logarithm of the data.}
	\label{fig:slow_measurement}
\end{figure*}

We show in Fig.~\ref{fig:slow_measurement} such comparison with a number of shots $N_S$ equal to 1024, 128, 64, 16, 4. Looking at the approximation ratio $R$ (Fig.~\ref{fig:slow_measurement}(a)) we see that taking $N_{S} = 128$ shots reduces $R$ by $5\%$ w.r.t. $N_{S} = 1024$ and going to $N_{S} = 64$ reduces it by a further $5\%$. This behaviour then stops and even reverses its trend. In fact, we even see an average increase going from $N_{S} = 16$ to $4$. This is understandable since the reconstruction of the energy with as little as 4 shots is not indicative of the real energy of the state. Specifically, from a final QAOA state we might sample the solution bitstring 2, 3 or even 4 times out of 4 and the expectation of the energy on these three samplings would be very different. This behaviour is indeed confirmed also by the fidelity in Fig.~\ref{fig:slow_measurement}(a) which follows the same trend as the approximation ratio. 

To have a better understanding of how the algorithm adapts to the sampling noise, we look at the kernel noise parameter $\sigma^2_N$ which is learned by the Gaussian process during the fitting at each step of the optimization (see Appendix~\ref{sec_BO_with_noise} for details on the noise hyperparameter). The plot in Fig.~\ref{fig:slow_measurement}(b) shows that, after an initial phase, the kernel noise sets at a specific value at around 400 steps. In addition to that, the lower the number of shots the larger the noise parameter learned. In fact, by fitting the average kernel noise found at the end of the training (Fig.~\ref{fig:slow_measurement}(c)) we obtain that $\sigma^2_N$ follows a power law with $N_S^{-1.1}$. This trend is comparable to the expected trend for the variance $N_S^{-1}$ of the reconstruction of the energy. This shows that the Gaussian process adapts to sampling noise.

\begin{figure}
	\centering
	\includegraphics[width = 0.48\linewidth]{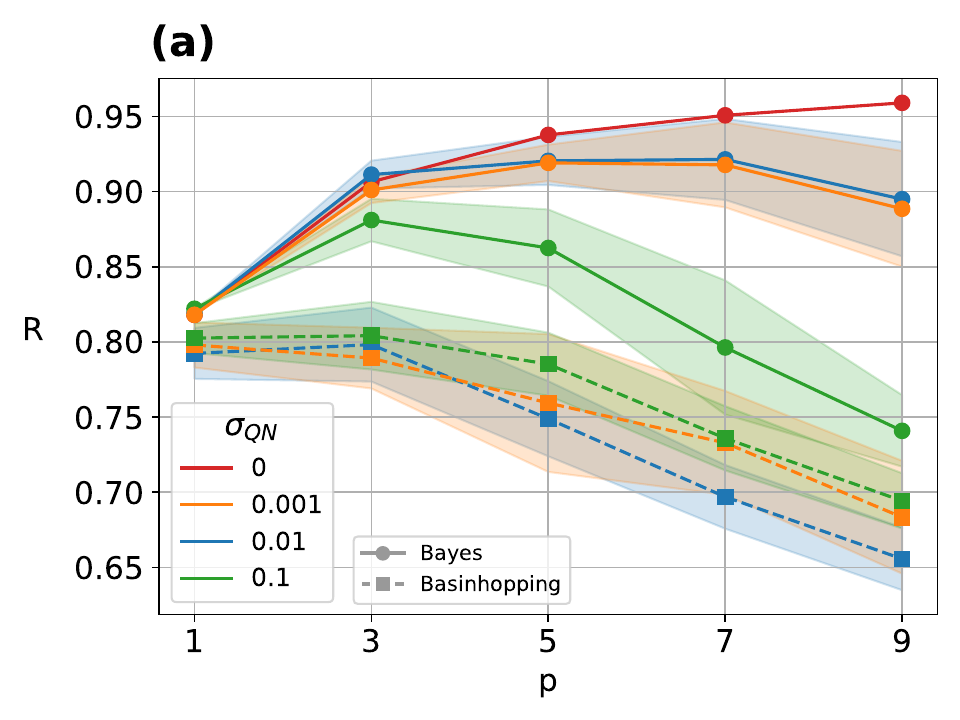}
	\includegraphics[width = 0.48\linewidth]{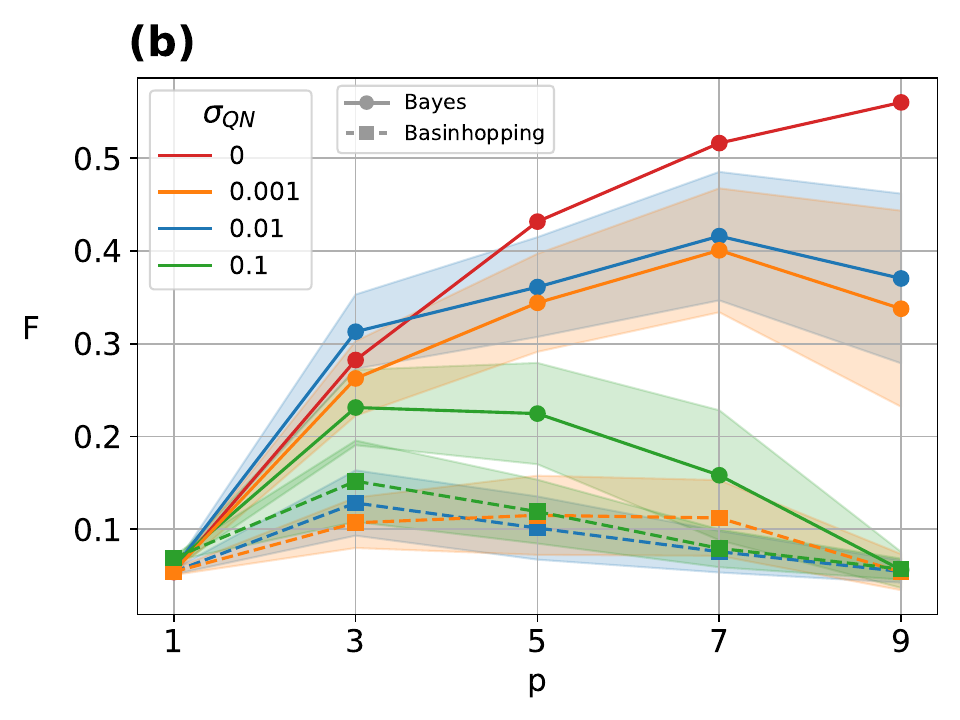}
	\caption{\textbf{Approximation ratio $R$ for different values of the quantum noise $\sigma_{QN}$.}  The noise is simulated by adding random Gaussian noise with mean zero and standard deviation $\sigma_{QN}$ to the variational parameters $(\boldsymbol{\gamma}, \boldsymbol{\beta})$. The plot on the left (right) shows the effects of the noise $\sigma_{QN}$ on the final obtained approximation ratio $R$ (fidelity $F$) as a function of the QAOA depth $p$ for different $\sigma_{QN}$. Shaded areas correspond to one standard deviation. The results are obtained on the graph with 10 nodes in Fig.~\ref{fig:graphs}(b).}
	\label{fig:noise}
\end{figure}

\textit{Quantum noise --}
Another relevant issue in the state-of-the-art NISQ devices are the sources of quantum noise which can interfere with the quantum circuit. Every device has different sources of noise depending on the underlying technology. In order to simulate it without specifying the device technology we add random noise on every QAOA parameter. In this way Eq.~\eqref{eq:qaoa_state} for the Max-Cut problem is modified as:
\begin{align}
\begin{split}
	|\boldsymbol{\theta}\rangle = \prod_{l=1}^p e^{-i \sum_{i} \beta_l^i X_i}\ e^{i\sum_{\langle i,j \rangle} \gamma_l^{(i,j)} Z_i Z_j} |+  \rangle,
\end{split}
\end{align}
where $\beta_l^i$ and $\gamma_l^{(i,j)}$ act differently on every qubit/edge of the graph at every layer because they are affected by Gaussian random noise with mean zero and standard deviation $\sigma_{QN}$. In Fig.~\ref{fig:noise} we plot $R$ and $F$ as a function of depth at different values of $\sigma_{QN}$ on the 10 nodes graph for the Max-Cut problem. 

By increasing $\sigma_{QN}$ and $p$, we expect to obtain a worse approximation ratio $R$ because the error accumulates along the circuit as the number of parameters grows. Indeed, as we can see in the figure, from $p \geq 5$ the obtained $R$ decreases w.r.t.~the noiseless case, decreasing even by $20\%$ for $p = 9$ with $\sigma_{QN} = 0.1$. Considering $\sigma_{QN} = 0.001, 0.01$, both $R$ and $F$ grow/remain stable up to $p = 7$ which indicates that, for shallow circuits, Bayesian optimization is robust against noise. 

We care to stress that the case $\sigma_{QN} = 0.1$ was considered in order to show the effect of an exponential growth of machine noise. Realistically, a Gaussian white noise with variance $0.1$ affecting each of the parameters (in range $[0, \pi]$, see Appendix~\ref{sec_details_BO_2}), would completely destroy the state preparation. In fact at $p = 9$ the fidelity $F$ is the same as $p=1$ (Fig.~\ref{fig:noise} (a)). 

We compare the results of our algorithm with the second best performing algorithm of Fig.~\ref{fig:comparison}, basin-hopping. It is clear that when subjected to noise this algorithm performs poorly. Considering the approximation ratio, basin-hopping shows barely any improvement with respect to the depth of the circuit with the results starting to plummet from $p=3$. Most importantly, the fidelity peaks at $p=3$ with $F \simeq 0.15$ (Fig.~\ref{fig:noise} (b)) and then remains contained under this value. The poor performance in terms of fidelity confirms that basin-hopping, while being an effective algorithm in the noise-free scenario, is not apt for optimization in the presence of noise. This is probably due to the fact that basin-hopping is a global optimizer that exploits a local gradient-based optimization routine (see Appendix~\ref{other_optimizers}). Calculating gradient in the presence of noise is in fact non-optimal since even a small variation of the parameters can impact greatly the evaluation of the function, returning a gradient that does not represent the local structure of the landscape.

\section{Conclusions}\label{Conclusions}
In this work we have presented the Bayesian optimization algorithm as a subroutine to optimize the variational parameters of the QAOA. We have applied it to find the solutions of two combinatorial optimization problems, the Max-Cut and the MIS on two graph instances.

After introducing the QAOA and the details of the Bayesian optimization algorithm, we have focused on its capability to adapt to the data and to predict new possible optimal points by exploiting both the accumulated knowledge from the previous observations and the uncertainty with respect to the optimization landscape.

We have analyzed some details of the QAOA at low circuit depth with the purpose of presenting some of the issues related to the optimization of a variational quantum algorithm. These include the flatness of the energy landscape and the limited information that we can retain from the energy of the QAOA state compared to its overlap with the ground state. After that, we have compared the Bayesian optimization with other global optimization methods, and we have showed that they require more calls, in the order of tens or hundreds, to the quantum circuit with respect to the Bayesian optimizer. This is a first sign that this method responds more efficiently to the requirements from the quantum part of the QAOA. 

Secondly, we have analyzed the effects of a finite number of measurements for the reconstruction of the energy landscape. We have showed that the results are slightly altered by a $5\%$ decrease in the approximation ratio by using 1024 measurements compared to the optimization with the exact energy. A lower number of measurements will results in a decreasing approximation ratio. We have also showed that the Gaussian process learns to add a noise hyperparameter which is proportional to the variance expected from the reconstruction of the energy. This can be seen as a further example of adaptation of the Bayesian algorithm to the data. 

Finally, we have simulated a noisy algorithm and we have showed that for shallow circuits, with depth $p \in [1,3, 5, 7]$, approximation ratio and fidelity are improved even for reasonable values of the noise. For deeper circuits, up to $p \geq 9$, the intensity of noise sensibly affects the final approximation ratio, as expected. Eventually, we compared it to basin-hopping, which uses a local gradient-based optimizer, and showed that it performs very poorly, suggesting that a gradient-free optimizer is indeed the better choice.

These findings show that Bayesian optimization is a robust method which can account for both quantum and sampling noise. For this reason it represents a valid tool for solving optimization problems via hybrid algorithms to be run on a NISQ device.

\section*{Code and data availability}
Code and data are available from the corresponding author, upon reasonable request.
\section*{Acknowledgments}
We thank C.~Sanavio and F.~Dell'Anna for useful discussions.
This research is funded by the International Foundation Big Data and Artificial Intelligence for Human Development (IFAB) through the project “Quantum Computing for Applications”.
E.~Ercolessi, S.~Tibaldi and D.~Vodola are partially supported by Istituto Nazionale di Fisica Nucleare (INFN) through the project “QUANTUM” and the QuantERA 2020 Project “QuantHEP”.

\bibliographystyle{unsrt}
\bibliography{biblio}

\appendix
\section{Details on the Bayesian optimization}\label{sec_details_BO}
In this appendix we give a general overview of the Bayesian terms used in the paper and a detailed review of the algorithm we used in this work to perform the Bayesian optimization for the QAOA.

\subsection{Overview of Bayesian terminology}\label{sec_details_BO_1}
The Gaussian process is a surrogate model which aims at reconstructing the landscape of optimization of an unknown function $f(\btheta)$. It is one of the two main ingredients of the Bayesian optimization algorithm, along with the acquisition function. We can now define the Bayesian terms used throughout the paper.
\begin{enumerate}
\item Prior $p(\boldsymbol{f})$. This distribution encapsulates the previous knowledge we have about the target function $f$. Typically, it consists in a multivariate normal distribution centered around 0 in which the covariance between points is assigned by a kernel function like Eq.\eqref{matern_kernel}:
\begin{equation}
	p(\boldsymbol{f}) = \mathcal{N} (\boldsymbol{0}, k(x, x')).
\end{equation}
To make an initial guess on our function we can sample a function $\boldsymbol{f}^\ast$ from this distribution. To do so, we choose a set of points $\bTheta$ and evaluate:
\begin{equation}
\label{eqn_sampling_prior}
	\boldsymbol{f}^\ast \sim \mathcal{N}(\boldsymbol{0}, k(\bTheta, \bTheta)).
\end{equation}

\item Likelihood $p(\boldsymbol{y} | \boldsymbol{f})$. This distribution represents the compatibility between the prior $p(\boldsymbol{f})$ and the observations $\boldsymbol{y}$. In the context of Gaussian processes it is defined by another Gaussian distribution.

\item Posterior $p(\boldsymbol{f} | \boldsymbol{y})$. This describes the knowledge we have about $f$ after having collected some observations $\boldsymbol{y} = f(\bTheta)$. The aim of the Gaussian process is to make the posterior generate functions as similar as possible to $f$. It is related to the prior and likelihood through the Bayesian theorem:
\begin{equation}
\label{eqn_bayes_theorem}
	p(\boldsymbol{f} | \boldsymbol{y}) = \frac{p(\boldsymbol{y} | \boldsymbol{f}) p(\boldsymbol{f})}{p(\boldsymbol{y})},
\end{equation}
which states that a posterior distribution is proportional to their product. In the context of Gaussian processes, the posterior is calculated from the prior by an operation which is called \textit{conditioning}, resulting in 
\begin{equation}
	\label{eqn_posterior}
	p(\boldsymbol{f} | \boldsymbol{y})  \sim \mathcal{N}(\mu', k').
\end{equation}
Where the new mean and covariance are defined in the text (Eqs.~\eqref{eqn_new_mean},~\eqref{eqn_new_cov}). 
From Eq.~\eqref{eqn_posterior} we see that the posterior is itself a Gaussian multivariate distribution. Therefore, we can sample functions $\boldsymbol{f}^\ast$ as we do with the prior~\eqref{eqn_sampling_prior} but now their values will coincide with $\boldsymbol{y}$ at every point $\bTheta$ where we sampled $f$.

\item Marginal likelihood $p(\boldsymbol{y})$. Also called \textit{evidence}, it is the normalization term in Eq.~\eqref{eqn_bayes_theorem}. In Bayesian inference it represents the total probability of generating the observed samples $\boldsymbol{f}$ from the prior. It is indeed obtained integrating over all possible function values $\boldsymbol{f}$:
\begin{equation}
	p(\boldsymbol{y}) = \int p(\boldsymbol{y} | \boldsymbol{f}) p(\boldsymbol{f}) d \boldsymbol{f}.
\end{equation}
Like the posterior distribution, it has a closed form in term of a multivariate normal distribution. Thus, the maximization of its logarithm is used in Gaussian processes to pick the best hyperparameters (as done in Sec.~\ref{sec_hyperparam}).
\end{enumerate}

\begin{figure}
	\centering
	\includegraphics[width = 0.9\linewidth]{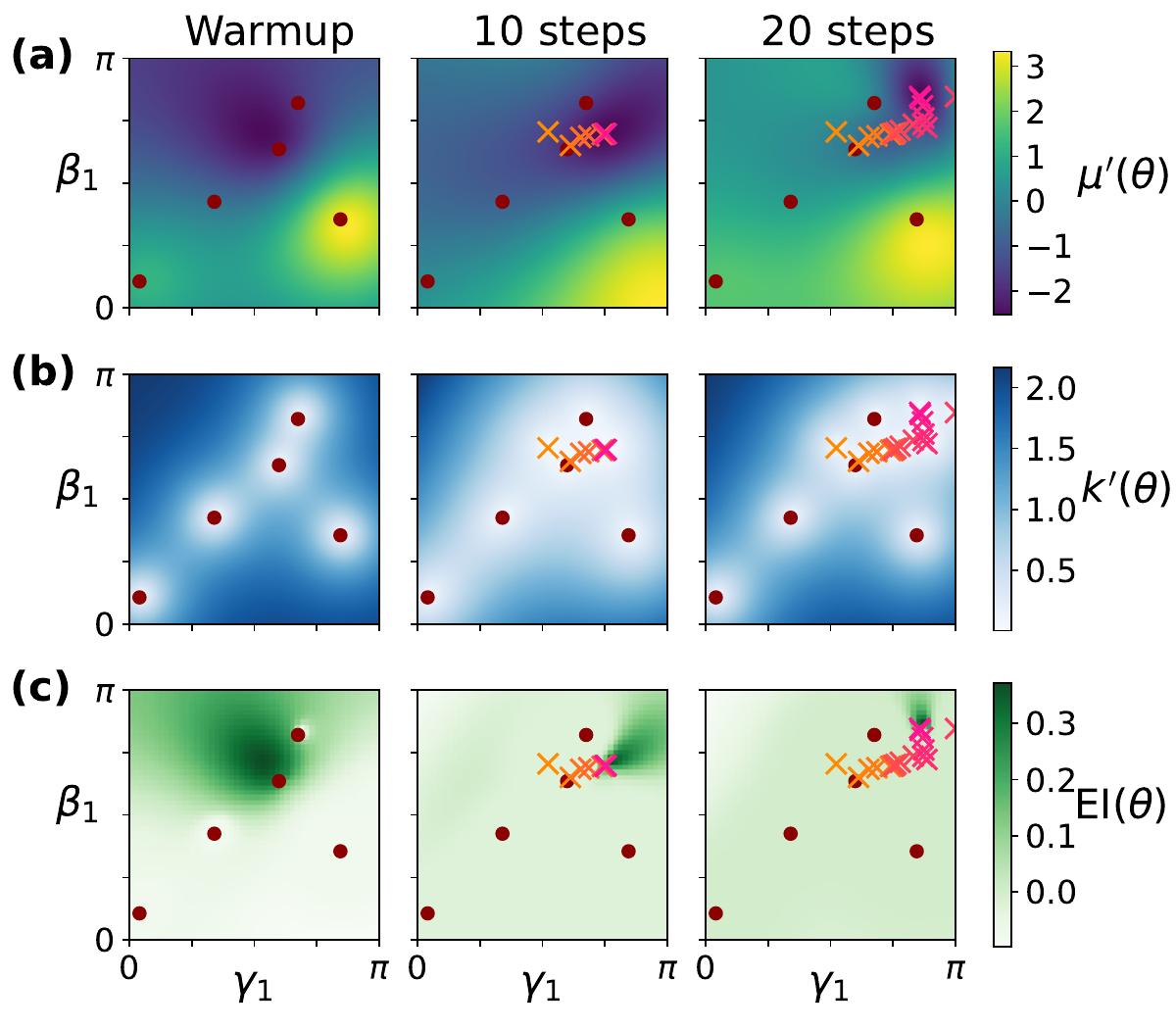}
	\caption{\textbf{Posterior and acquisition function during optimization.} Here we plot the posterior and the acquisition function at three different steps of the optimization: (from left to right) at warmup, after 10 steps and after 20 steps. In rows (a) and (b) we plot respectively the mean $\mu '(\boldsymbol{\theta})$ and  the variance $k '(\boldsymbol{\theta})$ of the posterior while in row (c) the acquisition function $\mathrm{EI}(\btheta)$ (see Eqs.~\eqref{eqn_new_mean} - \eqref{eqn_acquisition_function} in the text). These data, obtained running the Bayesian optimization on QAOA with $p=1$ on the graph of Fig.~\ref{fig:graphs}(a), are shown as a function of the two variational parameters $\btheta = (\gamma_1, \beta_1)$. The red dots indicate the warmup points ($N_W = 5$, in this case) while the crosses are the points that have been selected by subsequents Bayesian optimization steps as new candidate solutions (with a color scale from first (orange) to last (pink)). At every Bayesian optimization step, $\mu '(\boldsymbol{\theta})$ encodes the knowledge of the landscape of the function to optimize, while $k' (\boldsymbol{\theta})$ contains the uncertainty. The acquisition function $\mathrm{EI}(\btheta)$ combines the information from $\mu '$ and $k'$ and the position of its maximum proposes the next possible optimal point. From these considerations, we see that after 20 steps the mean $\mu '(\boldsymbol{\theta})$ (row (a)) recreates the landscape (our knowledge of $E(\btheta_1)$) with more precision than at warmup (compare with the real energy landscape in Fig.~\ref{fig:landscapes}(a)). In addition to that, $\mathrm{EI}(\btheta)$ becomes more and more flat in all the landscape except for a small area in the top right corner on the right panel of (c). This means that the Gaussian process has acquired enough knowledge from the data to converge to the energy minimum.}
	\label{fig:posterior}
\end{figure}

\subsection{Bayesian optimization algorithm}\label{sec_details_BO_2}
This algorithm is made out of three phases: warmup, kernel optimization, acquisition function maximization and is summarized in Algorithm~\ref{alg_pseudocode}.

\textit{Warmup --} In the warmup phase we start with a set of $N_{W} = 10$ points $X = \{\btheta_j\}_{j=1}^{N_W}$ with $\btheta_j \in \mathbb{R}^d$ where $d = 2p$ and $p$ is the depth of the QAOA circuit. Each point $\btheta_j$ is a set of angles $(\boldsymbol{\gamma}, \boldsymbol{\beta})$. These points are sampled from the latin hypercube of bounds $[0, \pi]^d$. For each set of angles we estimate the energy of QAOA $y_j$ and we create the design matrix $\bTheta = (\btheta_1,\dots,\btheta_{N_W})$ and the observation vector $\by$  with the energies of the points $\btheta_j$. We also store the point with lowest energy as $(\btheta_m, f_m)$. At this step there is no calculation involving the Gaussian process which is currently set to its prior: a multinomial gaussian distribution centered around zero.

\begin{figure*}
	\centering
	\includegraphics[width = \linewidth]{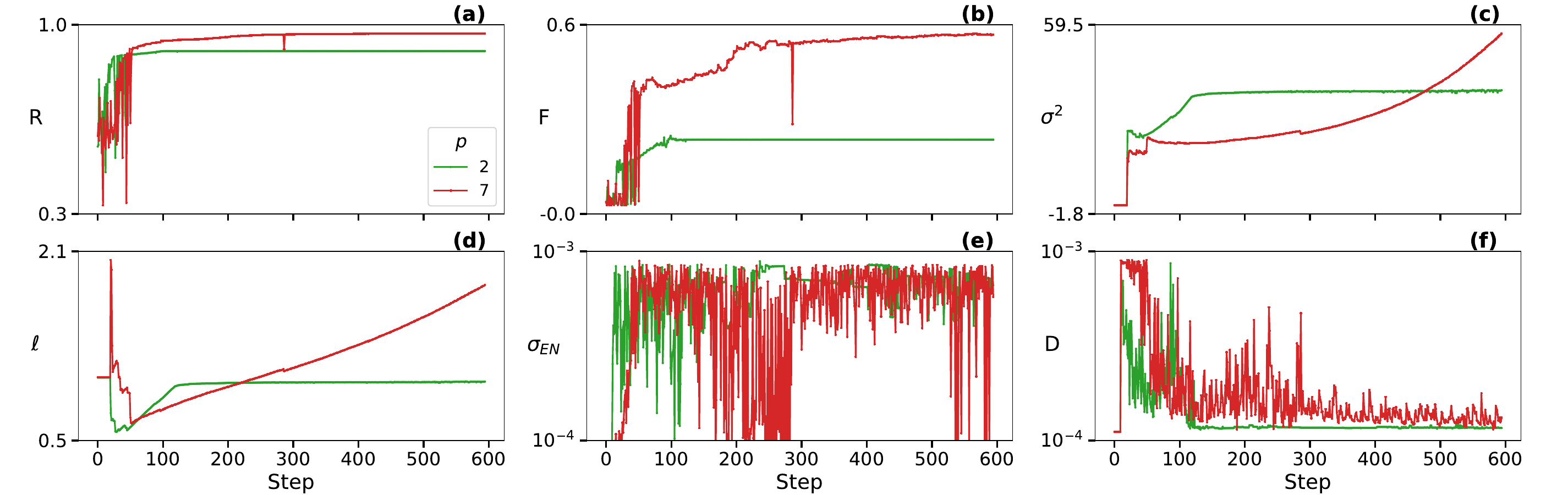}
	\caption{\textbf{Parameters of Bayesian optimization at p = 2, 7.} Plots of the parameters changing during Bayesian optimization for two runs with $N_{BAYES} = 600$ steps. (a) Approximation ratio $R$. (b) Fidelity $F$. (c) Kernel constant $\sigma^2$. (d) Kernel correlation length $\ell$. (e) Standard deviation of the expected improvement and (f) average distance of the points of differential evolution at the last generation $N_T$.}
	\label{fig:training}
\end{figure*}

\textit{Kernel optimization --} In this part we look for the hyperparameters $(\tilde{\sigma}, \tilde{\ell})$ which maximize the marginal likelihood function $p(\boldsymbol{y}|\bTheta)$ (Eq.~\eqref{eqn_log_marginal_likelihood}). This optimization is performed by repeating the \texttt{L-BFGS}~\cite{Liu1989} minimization on $-p(\boldsymbol{y}|\bTheta)$ for 10 times and selecting the best parameters found. The parameters found at every step of the optimization are plotted in panels (c) and (d) of Fig.~\ref{fig:training}.

\textit{Acquisition function maximization --} Once the hyperparameters are set the algorithm exploits its knowledge and uncertainty of the data to propose a new point $\btheta'$ with the new parameters where we evaluate the QAOA circuit. This is done by maximizing the expected improvement in Eq.~\eqref{eqn_acquisition_function}.

The new point $\btheta'$ maximizing the expected improvement is then added to the dataset $\mathcal{D}$ and the algorithm is repeated. The procedure stops after $N_\mathrm{BAYES}$ iterations. Fig.~\ref{fig:posterior} provides an illustrative example (with $p=1$, $N_{W} = 5$, and $N_\mathrm{BAYES}=20$) of how the Bayesian method operates and moves through the landscapes.

\section{Bayesian optimization with noise}\label{sec_BO_with_noise}
In this paper we consider two scenarios in which the energy $E(\btheta)$ is affected by a source of noise: the finite number of samplings and the quantum noise at the gate level. In the context of Bayesian optimization we can account for noise modifying the kernel function by adding a term $\sigma^2_N \mathbb{I}$ like so:
\begin{equation}
	k(\btheta, \btheta') = \sigma^2 \bigg( 1 + \frac{\sqrt{3} \norm{\btheta - \btheta'}_2}{\ell} \bigg) e^{-\frac{\sqrt{3} \norm{\btheta - \btheta'}_2} {\ell}} + \sigma_{N}^2 \mathbb{I}
\end{equation}

This is usually called a white kernel and it is a parameter added to the diagonal to account for random fluctuations around the true value of $f(\btheta)$. In this way the new predicted mean and covariance for a set of data $\bTheta$ can be easily calculated to be:
\begin{align}
	\mu' &= \bkappa^T \cdot (\bKappa + \sigma^2_{N} \mathbb{I})^{-1}\cdot\by \label{eqn_new_mean_noise}\\
	k' &= k(\btheta, \btheta) - \bkappa^T  \cdot (\bKappa + \sigma^2_{N} \mathbb{I})^{-1}\cdot \bkappa	\label{eqn_new_cov_noise}
\end{align}

The constant $\sigma_{N}^2$ belongs to the list of hyperparameters (along with $\sigma^2$ and $\ell$) that are optimized with the log marginal likelihood which now takes the form
\begin{align*}
	\log p(\boldsymbol{y} | \bTheta) = 
	 - &\frac{1}{2} \boldsymbol{y}^T \cdot (\bKappa + \sigma^2_{N} \mathbb{I})^{-1} \cdot \boldsymbol{y} \\
	 - &\frac{1}{2} \log \det(\bKappa + \sigma^2_{N} \mathbb{I}) \\
	 - &\frac{N}{2} \log 2 \pi.
	\label{eqn_log_marginal_likelihood_noise}
\end{align*}
We show how the hyperparameter $\sigma^2_N$ is learned during training in the text in Fig.~\ref{fig:slow_measurement}(b).

\section{Differential evolution}\label{sec_diff_ev}
Finding the point  $\tilde{\btheta}$ of the parameter space that maximizes the expected improvement $\mathrm{EI}(\btheta)$ (Eq.~\ref{eqn_acquisition_function}) is not an easy task since $\mathrm{EI}(\btheta)$ can show a fairly flat landscape~\cite{Snoek2012}, in particular after many optimization steps (for example, see Fig.~\ref{fig:posterior}(c)). 

To compute the maximum of $\mathrm{EI}(\btheta)$ in this work we use the differential evolution algorithm~\cite{Price2006}. This is an evolutionary method in which populations of points $\{ \btheta \}$, called generations, are iteratively obtained from the previous ones until convergence. The algorithm starts by initializing a generation (a) and then the population is updated following three main steps: (b) mutation, (c) cross-over and (d) selection.

(a) We choose as starting population $N_P = 15 \cdot 2p$ points $\{ \boldsymbol{\theta}_{i,1} \}$ where the index $i \in \{1,\dots, N_P\}$ uniquely identifies the point within the belonging population, while the index $1$ indicates that the point belongs to the first generation $g = 1$. These points are randomly generated on the latin $2p$-cube of bounds $[0, \pi]^{2p}$ and to each point there is an associated expected improvement $\mathrm{EI}(\boldsymbol{\theta}_{i,1})$.

(b) For each $\boldsymbol{\theta}_{i,g}$ (called parent point) in the population, the differential evolution picks three random points, different from $\boldsymbol{\theta}_{i,g}$, labelled by $r_0, r_1, r_2$ within the corresponding population, and creates a new point as:
\begin{equation}
	\boldsymbol{v}_{i,g} = \boldsymbol{\theta}_{r_0,g} + M(\boldsymbol{\theta}_{r_1,g} - \boldsymbol{\theta}_{r_2,g}),
	\label{eq:mutation}
\end{equation}
where $M \in (0.5,1)$ is a hyperparameter that is selected randomly at every generation. Through Eq.~\ref{eq:mutation} the differential evolution mutates and recombines the population to create another set of parent points $\boldsymbol{v}_{i,g}$.

(c) A new point $\boldsymbol{u}_{i,g}$ (offspring point) is created from $\boldsymbol{\theta}_{i,g} = (\theta_{i,g}^{1}, \dots, \theta_{i,g}^{2p})$ and $\boldsymbol{v}_{i,g} = (v_{i,g}^1, \dots, v_{i,g}^{2p}) $ choosing randomly between their coordinates $\theta_{i,g}^j$ and $v_{i,g}^j$ for every $j = 1, \dots, 2p$.

(d) Finally, if $\mathrm{EI}(\boldsymbol{u}_{i,g}) \geq \mathrm{EI}(\boldsymbol{\theta}_{i,g})$ the algorithm replaces $\boldsymbol{\theta}_{i,g}$ with $\boldsymbol{u}_{i,g}$ in the next generation, otherwise $\boldsymbol{\theta}_{i,g}$ is kept.

Steps (b) through (d) are repeated for $N_{T}$ generations $g$. The algorithm stops when two convergence criteria are fulfilled: the standard deviation $\sigma_{EN}$ of the population's expected improvement (see Fig.~\ref{fig:training}(e)) and the average distance $D$ among the population points (see Fig.~\ref{fig:training}(f)) are below a certain threshold (we set $\sigma_{EN} = D = 10^{-3}$).  When this happens, the point in the population with the maximum expected improvement is selected as $\tilde{\btheta}$. 
We notice that the criterion on the distance guarantees that in a flat landscape like the one of $\mathrm{EI}(\btheta)$ the points do not get stuck on a plateau and concentrate closer to a unique candidate maximum.
Although the algorithm requires many evaluations of $\mathrm{EI}(\boldsymbol{u}_{i,g})$ (as shown also in the main text) it is a valid algorithm for finding the maximum within the flat landscape of the acquisition function.

\section{Other optimizers}\label{other_optimizers}
\textit{Basin-hopping -- }Basin-hopping is a global stochastic optimization algorithm~\cite{Wales_1997}. It combines two steps: (i) a local optimization which proposes a candidate solution and (ii) a perturbation of such candidate in order to make it hop to other basins which might contain a global optimal point. The new point is accepted or rejected according to a probability which depends on a ``temperature'' parameter. The ``temperature'' parameter decreases with the iteration number so that, at the beginning, new proposals are easily accepted while, at larger iterations, the algorithms becomes more and more selective. The algorithm runs for a fixed number of iterations and the local optimizer used in this context is the gradient based $BFGS$ algorithm.

\textit{Dual annealing --} This global optimization algorithm is the generalized form of the simulated annealing and it is paired with a local optimization which is performed at the end of the annealing to refine the solution~\cite{XIANG1997216}. It is a variation of a hill climbing algorithm in which a solution is randomly perturbed and the new proposed point is accepted with a probability that depends on the difference in energy between the two points. This probability also depends on a ``temperature'' parameter that, like in the basin-hopping case, decreases with the number of iterations in order to converge to a candidate solution.

\end{document}